\documentclass[11pt]{article}
\usepackage{wds11,epsf}

\usepackage[square, numbers, comma, sort&compress]{natbib}
\usepackage{graphicx}
\usepackage{amsthm}
\usepackage{amsmath}
\usepackage{amsfonts}
\usepackage{mathtools}

\lefthead{NOSEK T.}
\righthead{EFFECTS OF MATTER IN NEUTRINO OSCILLATIONS \dots}

\begin{document}

\title{Effects of Matter in Neutrino Oscillations and Determination of Neutrino Mass Hierarchy at~Long-baseline Experiments}

\author{T. Nosek}

\affil{Institute of Particle and Nuclear Physics, Faculty  of  Mathematics  and  Physics,\linebreak Charles University, Prague, Czech Republic}

\begin{abstract}
Neutrino oscillations change when in media in comparison to vacuum oscillations due to the scattering of neutrinos on matter constituents, electrons particularly. This can be easily described by introducing new effective matter mixing angles and squared mass-splittings. Exploiting the matter effects and subsequent enhancement or suppression of oscillation probabilities can be used to determine the hierarchy of neutrino mass states. Recent long-baseline experiments NO$\nu$A and T2K investigate this possibility. Together NO$\nu$A and T2K combined can reject the wrong hierarchy for more than 20\% of all possible values of yet unknown CP violating phase $\delta \in [-180^\circ, 180^\circ]$.
\end{abstract}

\begin{article}

\section{Introduction}
Almost all of the neutrino oscillation parameters are already known with good precisions~\cite{PDGrev}. Mixing angle $\sin^2 \theta_{12} = 0.308 \pm 0.017$ and squared mass-splitting $\Delta m_{21}^2 = 7.54^{+0.26}_{-0.22} \times 10^{-5}$~eV$^2$ are known from oscillations of solar neutrinos and experiment KamLAND (see e.g. \cite{superk:solarresults2016,kamland:results2013}). The size of mixing angle $\sin^2 \theta_{23} \approx 0.5$ and the absolute value of the larger mass-splitting $\vert \Delta m^2 \vert = (2.4 \pm 0.1) \times 10^{-3}$~eV$^2$ were measured in oscillations of atmospheric neutrinos and long-baseline experiments \citep{minos:results2013, t2k:results2014, nova:numuresults2016}. Reactor experiments Daya Bay \cite{dayabay:results2016}, RENO \cite{reno:results2015} and Double Chooz \cite{doublechooz:results2015} gave us the last mixing angle $\sin^2 \theta_{13} = 0.024 \pm 0.002$. 
\par In standard 3$\nu$-model of oscillations, there is few left to answer. Do oscillations violate CP symmetry, what is the size of $\delta$ phase? Is 23-mixing maximal, i.e.~$\theta_{23} = 45^\circ$? If not, is it $>$ or $<$~$45^\circ$? What is the mass hierarchy, i.e.~what is the sign of $\Delta m^2$? Is there normal (NH; $\Delta m^2 > 0$) or inverted (IH; $\Delta m^2 < 0$) hierarchy?
\par A simple way to resolve the last question and to determine the mass hierarchy is to take advantage of matter effects on neutrino oscillations and study their incidental modifications in~comparison to vacuum oscillations. Recent long-baseline experiments NO$\nu$A and T2K measure oscillation probabilities $P(\nu_\mu \rightarrow \nu_e)$ and $P(\bar{\nu}_\mu \rightarrow \bar{\nu}_e)$. The current status of oscillation parameters makes it possible for these experiments to address the problem of mass hierarchy.
\par The text concerns itself with the basics of matter effects in $3\nu$-model of oscillations and the possibility of neutrino mass hierarchy determination using $\nu_\mu \rightarrow \nu_e$ channel. The predicted sensitivities of NO$\nu$A and T2K to determine the mass hierarchy are shown.
\section{CC Scattering of Neutrinos and Effective Neutrino Oscillation Parameters}
When propagating through media, neutrinos undergo weak interactions with other particles via exchange of $W^\pm$ (CC) and $Z^0$ gauge bosons (NC). Assuming that matter consists only of $p$, $n$ and $e^-$, all neutrino flavors are treated as equal in NC interactions, while only $\nu_e$ interacts with $e^-$ through CC \cite{wolfenstein:1977, smirnov:matter}. This intrinsic assymetry among the neutrino flavors in matter causes a distortion of vacuum oscillation patterns. The way to describe this is to introduce new effective matter oscillation parameters: $\theta_{ij} \rightarrow \Theta_{ij}, \Delta m_{ij}^2 \rightarrow \Delta M_{ij}^2$~\cite{smirnov:matter}.
\par For the purpose of oscillations only interaction potential $V$ corresponding to CC interactions can be added to the original vacuum Hamiltonian $H$ to form new matter effective $\mathcal{H}$. $V = \sqrt{2}G_FN_e$, with Fermi coupling constant $G_F$ and electron density $N_e$. The potential for antineutrinos can be formally obtained as $V \rightarrow -V$. Matter Hamiltonian $\mathcal{H}_f$ in the basis of~neutrino flavors $\nu_f \equiv (\nu_e, \nu_\mu, \nu_\tau)^T$ has a simple form
\begin{equation}
\label{EqMatterEffHf}
\mathcal{H}_f = \dfrac{\Delta m^2}{2E} \left[ U\begin{pmatrix}
0	&	& \\
	& a & \\
	& 	& 1\\
\end{pmatrix}U^\dagger + \begin{pmatrix}
\frac{2VE}{\Delta m^2}	&	& \\
	& 0	& \\
	&	& 0 \\
\end{pmatrix}\right],
\end{equation}
where $a \equiv \Delta m_{21}^2/\Delta m^2$ and $U \equiv U_{23}I_\delta U_{13}U_{21}$ is standard mixing matrix for three neutrino flavors \cite{kayser:NOphysics} as a product of three matrices $U_{ij} \equiv U(\theta_{ij})$ of rotation in the space of neutrino mass eigenstates and the matrix $I_\delta \equiv \mathrm{diag}(1,1,\exp(\mathrm{i}\delta))$ containing CP phase $\delta$, $E$ is the energy of $\nu$.
\par Neutrino mixing is defined with respect to the eigenstates of the Hamiltonian. In vacuum, the eigenstates are standard mass eigenstates $\vert \nu_i \rangle$ connected with flavor states $\vert \nu_\alpha \rangle$ via mixing matrix $U$. $U$ is a transformation matrix between the bases of mass states $\nu_m \equiv (\nu_1, \nu_2, \nu_3)^T$ and flavor states $\nu_f$ through relation $\nu_f = U\nu_m$. In matter, there are new effective matter mass eigenstates $\vert N_i \rangle$ with effective masses $M_i$ satisfying the eigenequation $\mathcal{H}\vert N_i \rangle = \dfrac{M_i^2}{2E}\vert N_i \rangle$. The new effective transformation matrix between bases $\nu_f$ and $N \equiv (N_1,N_2,N_3)^T$ is $\mathcal{U}$, i.e. $\nu_f = \mathcal{U}N$ and for Hamiltonian $\mathcal{H}_m$ in basis $N$ holds $\mathcal{H}_f = \mathcal{UH}_m\mathcal{U}^\dagger$.
\par One can find $\mathcal{U}$ through diagonalization of $\mathcal{H}_f$ and obtain new effective matter mixing angles $\Theta_{ij}$, and $M_i^2$, i.e.~new effective squared-mass splittings  $\Delta M_{ij}^2 = M_i^2 - M_j^2$. One simple way is to employ the smallness of parameters $a \approx 0.03$ and $\theta_{13}$ and receive approximative matter parameters, for details see \cite{smirnov:matter}. Then $\mathcal{U} \approx U_{23}(\theta_{23})I_\delta U_{13}(\Theta_{13})U_{12}(\Theta_{12})$, where $U_{ij}$ are again matrices of rotation in corresponding planes by the set angles. Due to the primary approximation $\theta_{23}$ can be taken as in vacuum. The other two angles are given by
\begin{equation}
\label{EqTh13}
\tan 2\Theta_{13} = \dfrac{\sin 2\theta_{13}}{\cos 2\theta_{13} - \frac{2EV}{\Delta m^2 (1-a\sin^2 \theta_{12})}} 
\end{equation}
and
\begin{equation}
\label{EqTh12}
\tan 2\Theta_{12} = \dfrac{a \sin2\theta_{12} \cos(\Theta_{13}-\theta_{13})}{a[\cos^2\theta_{12}-\sin^2\theta_{12}\cos^2(\Theta_{13}+\theta_{13})]-\sin^2(\Theta_{13}+\theta_{13})-\frac{2EV}{\Delta m^2}\cos^2 \Theta_{13}}.
\end{equation}
Fig.~1 shows $\sin^2 2\Theta_{12}$ and $\sin^2 2\Theta_{13}$ with respect to $VE$ in case of neutrinos/antineutrinos and NH/IH. Rather complicated formulae for $M_i^2$ are left out. Their dependence on $VE$ is depicted in Fig.~2. Note, that only difference $\Delta M_{ij}^2$ has a physical meaning, the absolute value of $M_i^2$ has not.
\par The oscillation probability in medium with constant and isotropic density such as in the case of long-baseline experiments has the same form as in vacuum \citep{smirnov:matter}. The elements of mixing matrix $U_{\alpha i}$ and squared mass-splittings $\Delta m_{ij}^2$ has to be traded for new ones $\mathcal{U}_{\alpha i}$ and $\Delta M_{ij}^2$, i.e.
\begin{equation}
P(\nu_\alpha \rightarrow \nu_\beta; L, E) = \sum_{i = 1} ^3 \sum_{j=1}^3 \mathcal{U}_{\alpha i}^* \mathcal{U}_{\alpha j}^{} \mathcal{U}_{\beta i}^{} \mathcal{U}_{\beta j}^* \exp \left( -\mathrm{i} \dfrac{\Delta M_{ij}^2}{2E} L\right),
\end{equation}
where $L$ is the oscillation baseline. Unlike in vacuum, the elements of effective mixing matrix in matter $\mathcal{U}_{\alpha i}$ are energy dependent and the oscillation length is changed thanks to $\Delta M_{ij}^2$.

\begin{figure}[t!]
\label{FigMatterAngles}
\begin{center}
\includegraphics[scale=0.8]{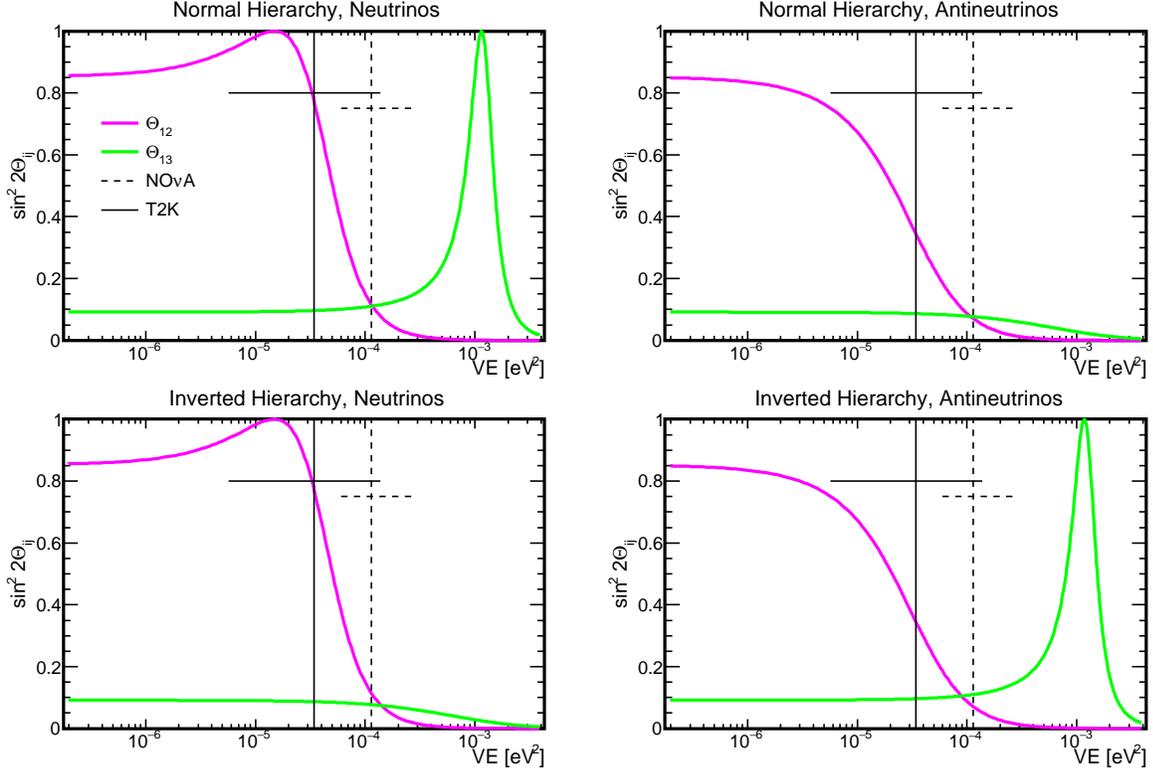}
\end{center}
\caption{The dependence of effective matter mixing angles $\sin^2 2\Theta_{ij}$ on the product of potential and neutrino energy $VE$. Corresponding range of neutrino energy spectra (horizontal lines) and their peaks (vertical lines) for NO$\nu$A (dashed, $E = 2.0$ GeV) and T2K (full, $E = 0.6$~GeV) were marked out, $\frac{G_F N_e}{\sqrt{2}} = \frac{1}{3500~\mathrm{km}}$. \textbf{Left semi-plane:} For neutrinos. \textbf{Right semi-plane:} For~antineutrinos. \textbf{Upper semi-plane:} For NH. \textbf{Lower semi-plane:} For IH.
}
\end{figure}

\begin{figure}[htb]
\label{FigMatterMasses}
  \begin{center}
    \includegraphics[scale=0.7]{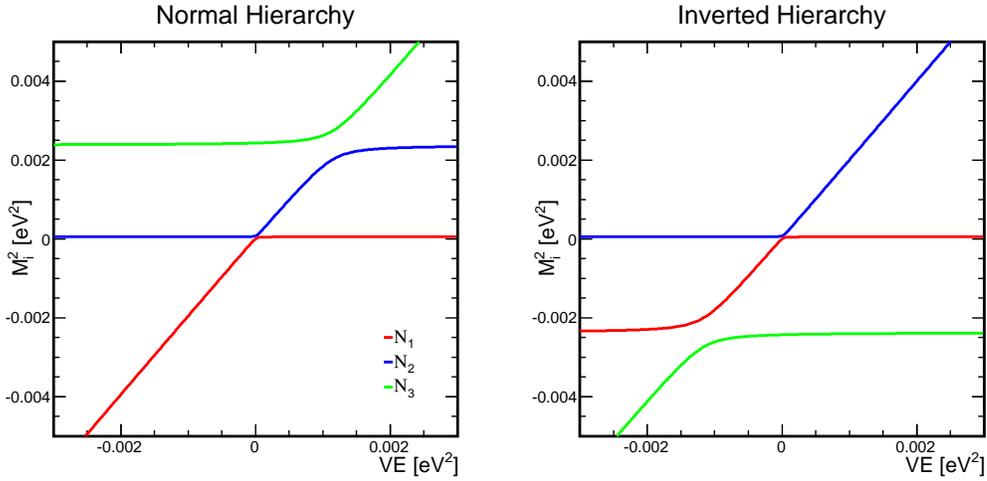}
  \end{center}
  \caption{The energy level scheme. Effective eigenvalues $M_i ^2$ of Hamiltonian in matter $\mathcal{H}$ as~functions of~$VE$. Note that for $VE \rightarrow 0$ $\Delta m_{21}^2$ and $\Delta m^2$ are recovered. \textbf{Left:} For NH. \textbf{Right:} For IH. Denotation of the effective mass eigenstates ($N_1, N_2, N_3$) respects the standard convention, i.e. $M_1^2 < M_2^2 < M_3^2$ in case of NH and $M_3^2 < M_1^2 < M_2^2$ in~case of IH. $VE < 0$ reflects the antineutrino case.}
\end{figure}

\section{Neutrino Mass Hierarchy Determination in $\nu_\mu \rightarrow \nu_e$ Channel}
According to Fig.~1 neutrino mass hierarchy can be determined by locating 13-resonance, i.e. $\sin^2 2\Theta_{13} \rightarrow 1$. The promising way is to study the $\nu_\mu \rightarrow \nu_e$ appearance channel on the scale of $L/E \approx 500$ km/GeV and subsequent enhancement or suppression of oscillation probability due to the effective increase or decrease of $\sin^2 2\Theta_{13}$. This is being done by two recent long-baseline experiments NO$\nu$A and T2K \cite{feldman:longbase} (range of their neutrino energy spectra with peaks can be also seen in Fig.~1) and is one of the main goals of the physical programme at future DUNE.

\par Unfortunately, it is difficult to reach the resonance itself in terrestrial conditions (higher energies and longer oscillation baselines are required). The ``relatively'' small matter effects provided by recent long-baseline experiments ($VE$ values, i.e.~neutrino energy spectra, for NO$\nu$A and T2K can be seen in Fig. 1) can be misidentified with uncertainties of some of the~measured or unknown oscillation parameters, particularly $\theta_{13}$, $\theta_{23}$ and $\delta$. $P(\nu_\alpha \rightarrow \nu_\beta)$ depends mainly on $\theta_{13}$, the octant of $\theta_{23}$ ($\theta_{23} >$ or $< 45^\circ$) and the (practically arbitrary) value of~$\delta$. The~$\theta_{13}$- and $\theta_{23}$-degeneracy of mass hierarchy determination at recent long-baseline experiments\linebreak are effectively eliminated with additional measurement of $\bar{\nu}$ oscillations. On the contrary,\linebreak $\delta$-degeneracy is always inherent. Therefore, in order to resolve the question of mass hierarchy using long-baseline experiments it is vital to measure both $\nu_\mu \rightarrow \nu_e$ and $\bar{\nu}_\mu \rightarrow \bar{\nu}_e$ oscillations.

\par The possible arrangements allowed by the odds -- NH, IH, upper hyper-plane (UHP) $\delta \in [0^\circ, 180^\circ]$, lower hyper-plane (LHP) $\delta \in [-180^\circ, 0^\circ]$, lower octant (LO) $\sin^2 \theta_{23} < 0.5$ and higher octant (HO) $\sin^2 \theta_{23} > 0.5$ -- are divided into two groups of favorable and unfavorable situations depending on the size of $\delta$. Hierarchy cannot be determined by recent experiments (T2K, NO$\nu$A) in unfavorable ones, i.e.~\textbf{NH-UHP-LO}, \textbf{NH-UHP-HO}, \textbf{IH-LHP-LO} and \textbf{IH-LHP-HP}. Favorable ones are \textbf{NH-LHP-HO}, \textbf{NH-LHP-LO}, \textbf{IH-UHP-HO} and \textbf{IH-UHP-LO}. In extremely favorable conditions ($\delta \approx 90^\circ, -90^\circ$) and NH-LHP-HO or IH-UHP-LO hierarchy can be determined even from $\nu$ oscillations only, in case of NH-LHP-NO or IH-UHP-HO even from $\bar{\nu}$ oscillations \cite{prakash:earlyantinu}.

\begin{figure}[t!]
  \begin{center}
    \includegraphics[scale=0.7]{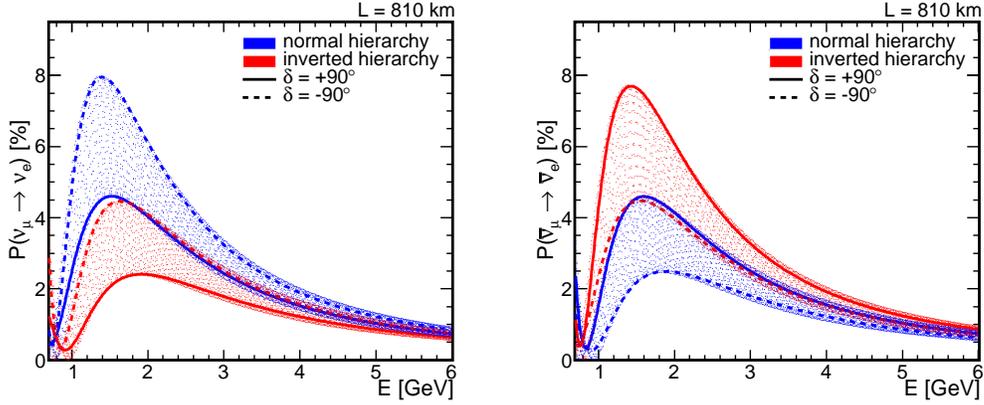}
  \end{center}
  \caption{Appearance probability of $\nu_e$ for initial $\nu_\mu$ as a function of energy $E$, $\frac{G_F N_e}{\sqrt{2}} = \frac{1}{3500~\mathrm{km}}$, $\theta_{23} = 45^\circ$, $\delta \in [-180^\circ, 180^\circ]$, $L = 810$ km. \textbf{Left:} For neutrinos. \textbf{Right:} For antineutrinos. 
  }
\end{figure}

\begin{figure}[t!]
	\begin{center}
		\includegraphics[scale=0.58]{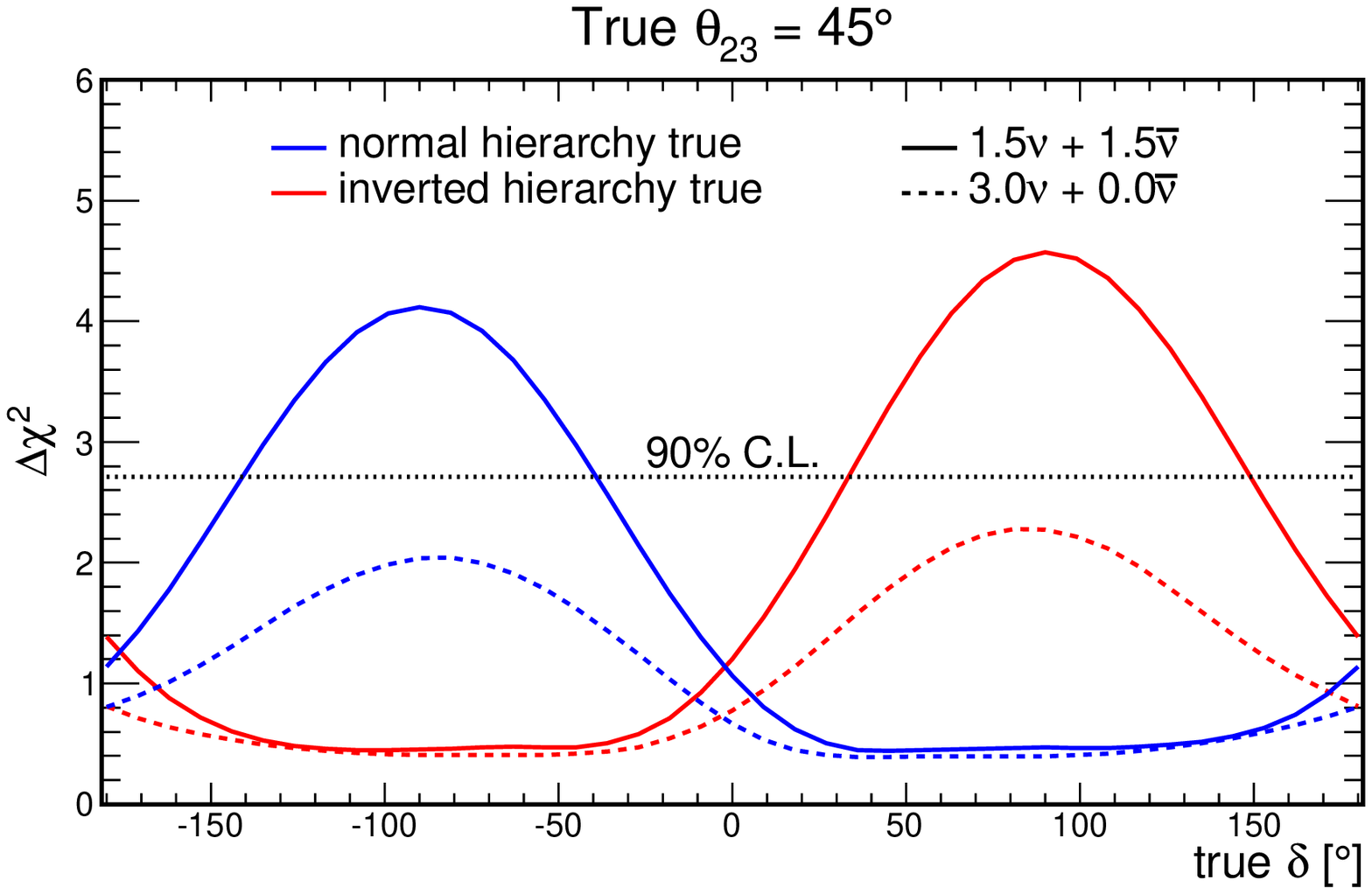}
	\end{center}
	\caption{$\Delta\chi^2$ (minimal $\chi^2$) for expected hierarchy resolution at NO$\nu$A with 3 years $\nu$ run (dashed) and 1.5y $\nu$ + 1.5y $\bar{\nu}$ (full) for true $\sin^2 \theta_{23} = 0.5$ assuming 10\% uncertainty in $\theta_{13}$, 6\% in $\Delta m^2$ with central values $\sin^2 \theta_{13} = 0.023$ (NH), 0.024 (IH) and $\vert \Delta m^2 \vert = 2.4 \times 10^{-3}$~eV$^2$. Marginalized over 90\% C.L. interval of $\sin^2 \theta_{23} \in [0.35, 0.65]$ and $\delta \in [-180^\circ , 180^\circ]$.}
\end{figure}

\begin{figure}[t!]
	\begin{center}
		\includegraphics[scale=0.58]{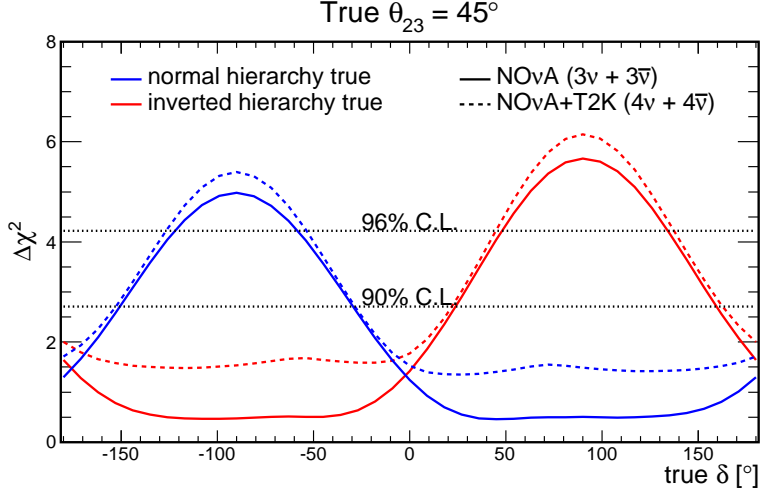}
	\end{center}
	\caption{$\Delta\chi^2$ (minimal $\chi^2$) for expected hierarchy resolution at NO$\nu$A solo (full) and combined with T2K (dashed) for true $\sin^2 \theta_{23} = 0.5$ assuming 10\% uncertainty in $\theta_{13}$, 6\% in $\Delta m^2$ with central values $\sin^2 \theta_{13} = 0.023$ (NH), 0.024 (IH) and $\vert \Delta m^2 \vert = 2.4 \times 10^{-3}$ eV$^2$ . Marginalized over 90\% C.L. interval of $\sin^2 \theta_{23} \in [0.35, 0.65]$ and $\delta \in [-180^\circ , 180^\circ]$.}
\end{figure}

\section{Neutrino Mass Hierarchy Determination at NO$\nu$A and T2K}
NO$\nu$A and T2K are both long-baseline oscillation experiments standing off-axis of their neutrino beams to tune the desired energy of $\nu$ and make their spectra much sharper with narrow width \cite{feldman:longbase}. They study $\nu_\mu$ disappearance and $\nu_e$ appearance in $\nu_\mu$ beams created in accelerator facilities of Fermilab (NO$\nu$A) and J-PARC (T2K), which can be switched to $\bar{\nu}$ mode to investigate $\bar{\nu}_\mu \rightarrow \bar{\nu}_e$ oscillations. Oscillation baseline is 810~km and $\nu$ energy ca 2 GeV in case of NO$\nu$A and 295 km and ca 0.6 GeV in case of T2K. NO$\nu$A is planned to run for 6 years, 3 in each of $\nu$, $\bar{\nu}$ modes and it started taking data in early 2014. The first analysis of $\nu_e$ appearance has been already published \cite{nova:nuefirstresults}. T2K began in 2009 with $\nu$ mode and $\bar{\nu}$ run started in 2015 with a plan of at least 4 years of $\nu$ and 4 years of $\bar{\nu}$. Fig.~3 shows the dependence of $P(\nu_\mu \rightarrow \nu_e)$ on $E$ at NO$\nu$A, $\delta$-degeneracy region commented earlier is clearly visible.

\par The possibility of neutrino mass hierarchy determination at NO$\nu$A and T2K was studied using \textsf{GLoBES} software \cite{globes:sims, globes:new3.0} with adjusted experiment definitions to be in agreement with their current status (source powers, target detector masses, fluxes etc.). The oscillation parameters values and uncertainties were taken as follows: $\sin^2 \theta_{13} = 0.023$ (NH), 0.024 (IH) with 10\% error, $\Delta m^2 = 2.4 \times 10^{-3}$ eV$^2$ with 6\% error, $\sin^2 \theta_{23} \in [0.35, 0.65]$ with true value 0.5, $\delta \in [-180^\circ, 180^\circ]$. The density of Earth's crust $\rho = 2.8$ g/cm$^3$, $\sin^2 \theta_{12} = 0.308$ and $\Delta m_{21}^2~=~7.54~\times~10^{-5}$~eV$^2$ were kept fixed.
\par The results are depicted in Fig.~4 and 5. Fig.~4 shows the importance of $\nu + \bar{\nu}$ data combination to determine neutrino mass hierarchy as was discussed in preceding section. The NO$\nu$A sensitivity would be improved even within a shorter period of data taking, concretely 3 years of $\nu$ data can be divided into 1.5 of $\nu$ and 1.5 of $\bar{\nu}$. The total sensitivity for NO$\nu$A only and combined with T2K after 6 and 8 years of running respectively is in Fig.~5. The favorable and unfavorable scenarios of hierarchy and $\delta$ can be very easily recognized. The wrong neutrino mass hierachy is rejected at 96\% C.L. by recent long-baseline experiments for ca 20\% of the possible values of $\delta \in [-180^\circ, 180^\circ]$.

\section{Conclusion}
The effects of matter in neutrino oscillations using effective matter parameters approach were described. The way to determine neutrino mass hierarchy in $\nu_\mu \rightarrow \nu_e$ channel at recent long-baseline experiments was discussed. In order to resolve neutrino mass hierarchy problem\linebreak both data on $\nu$ and $\bar{\nu}$ oscillations are needed because of inherent parameters correlations,\linebreak $\theta_{13}$ and $\theta_{23}$ in particular. According to the analyzed sensitivity studies NO$\nu$A and T2K can reject the wrong hierarchy at 96\% C.L. for more than 20\% of the whole interval of possible $\delta \in [-180^\circ, 180^\circ]$.

\acknowledgments 
{The study was supported by the Charles University in Prague, project\linebreak GA UK No. 362715.}


\bibliographystyle{egs}
\bibliography{references.bib}

\begin{thebibliography}{17}
\expandafter\ifx\csname natexlab\endcsname\relax\def\natexlab#1{#1}\fi
\expandafter\ifx\csname url\endcsname\relax
  \def\url#1{{\tt #1}}\fi
\expandafter\ifx\csname urlprefix\endcsname\relax\def\urlprefix{URL }\fi

\bibitem[{Abe et~al.(2014)}]{t2k:results2014}
Abe, K. et~al. [T2K Collab.], {Precise Measurement of the Neutrino Mixing Parameter
  $\theta_{23}$ from Muon Neutrino Disappearance in an Off-Axis Beam}, {\em
  Phys. Rev. Lett.\/}, {\em 112\/}, 181\,801, 2014, arXiv:1403.1532.\vspace*{-3mm}
\bibitem[{Abe et~al.(2016{\natexlab{a}})}]{superk:solarresults2016}
Abe, K. et~al. [Super-Kamiokande Collab.], {Solar Neutrino Measurements in Super-Kamiokande-IV}, {\em
  Phys. Rev.\/}, {\em D94\/}, 052\,010, 2016{\natexlab{a}}, arXiv:1606.07538.
\vspace*{-3mm}
\bibitem[{Abe et~al.(2016{\natexlab{b}})}]{doublechooz:results2015}
Abe, Y. et~al. [Double Chooz Collab.], {Measurement of $\theta_{13}$ in Double Chooz using neutron
  captures on hydrogen with novel background rejection techniques}, {\em
  JHEP\/}, {\em 01\/}, 163, 2016{\natexlab{b}}, arXiv:1510.08937.
\vspace*{-3mm}
\bibitem[{Adamson et~al.(2013)}]{minos:results2013}
Adamson, P. et~al. [MINOS Collab.], {Measurement of Neutrino and Antineutrino Oscillations
  Using Beam and Atmospheric Data in MINOS}, {\em Phys. Rev. Lett.\/}, {\em
  110\/}, 251\,801, 2013, arXiv:1304.6335.
\vspace*{-3mm}
\bibitem[{Adamson et~al.(2016{\natexlab{a}})}]{nova:nuefirstresults}
Adamson, P. et~al. [NOvA Collab.], {First measurement of electron neutrino appearance in
  NOvA}, {\em Phys. Rev. Lett.\/}, {\em 116\/}, 151\,806, 2016{\natexlab{a}}, arXiv:1601.05022.
\vspace*{-3mm}
\bibitem[{Adamson et~al.(2016{\natexlab{b}})}]{nova:numuresults2016}
Adamson, P. et~al. [NOvA Collab.], {First measurement of muon-neutrino disappearance in NOvA},
  {\em Phys. Rev.\/}, {\em D93\/}, 051\,104, 2016{\natexlab{b}}, arXiv:1601.05037.
\vspace*{-3mm}
\bibitem[{An et~al.(2016)}]{dayabay:results2016}
An, F.~P. et~al. [Daya Bay Collab.], {New measurement of $\theta_{13}$ via neutron capture on
  hydrogen at Daya Bay}, {\em Phys. Rev.\/}, {\em D93\/}, 072\,011, 2016, arXiv:1603.03549.
\vspace*{-3mm}
\bibitem[{Blennow and Smirnov(2013)}]{smirnov:matter}
Blennow, M. and Smirnov, A.~Y., {Neutrino propagation in matter}, {\em Adv.High
  Energy Phys.\/}, {\em 2013\/}, 972\,485, 2013, arXiv:1306.2903.
\vspace*{-3mm}
\bibitem[{Choi et~al.(2016)}]{reno:results2015}
Choi, J.~H. et~al. [RENO Collab.], {Observation of Energy and Baseline Dependent Reactor
  Antineutrino Disappearance in the RENO Experiment}, {\em Phys. Rev. Lett.\/},
  {\em 116\/}, 211\,801, 2016, arXiv:1511.05849.
\vspace*{-3mm}
\bibitem[{Feldman et~al.(2013)Feldman, Hartnell, and
  Kobayashi}]{feldman:longbase}
Feldman, G.~J., Hartnell, J., and Kobayashi, T., {Long-Baseline Neutrino
  Experiments}, {\em Advances in High Energy Physics\/}, {\em 2013\/},
  \urlprefix\url{http://dx.doi.org/10.1155/2013/475749}, 2013, arXiv:1307.7335.
\vspace*{-3mm}
\bibitem[{Gando et~al.(2013)}]{kamland:results2013}
Gando, A. et~al. [KamLAND Collab.], {Reactor On-Off Antineutrino Measurement with KamLAND}, {\em
  Phys. Rev.\/}, {\em D88\/}, 033\,001, 2013, arXiv:1303.4667.
\vspace*{-3mm}
\bibitem[{Huber et~al.(2005)Huber, Lindner, and Winter}]{globes:sims}
Huber, P., Lindner, M., and Winter, W., {Simulation of long-baseline neutrino
  oscillation experiments with \textsf{GLoBES}}, {\em Comput.Phys.Commun.\/},
  {\em 167\/}, 195, 2005, hep-ph/0407333.
\vspace*{-3mm}
\bibitem[{Huber et~al.(2007)Huber, Kopp, Lindner, Rolinec, and
  Winter}]{globes:new3.0}
Huber, P., Kopp, J., Lindner, M., Rolinec, M., and Winter, W., {New features in
  the simulation of neutrino oscillation experiments with GLoBES 3.0: General
  Long Baseline Experiment Simulator}, {\em Comput.Phys.Commun.\/}, {\em
  177\/}, 432--438, 2007, hep-ph/0701187.
\vspace*{-3mm}
\bibitem[{Kayser(2012)}]{kayser:NOphysics}
Kayser, B., {Neutrino Oscillation Physics}, 2012, arXiv:1206.4325.
\vspace*{-3mm}
\bibitem[{Olive et~al.(2014)}]{PDGrev}
Olive, K. et~al., {Review of Particle Physics}, {\em Chin.Phys.\/}, {\em
  C38\/}, 090\,001, 2014.
\vspace*{-3mm}
\bibitem[{Prakash et~al.(2014)Prakash, Rahaman, and
  Sankar}]{prakash:earlyantinu}
Prakash, S., Rahaman, U., and Sankar, S.~U., {The need for an early
  anti-neutrino run of NO$\nu$A}, {\em JHEP\/}, {\em 1407\/}, 070, 2014, arXiv:1306.4125.
\vspace*{-3mm}
\bibitem[{Wolfenstein(1978)}]{wolfenstein:1977}
Wolfenstein, L., {Neutrino Oscillations in Matter}, {\em Phys. Rev.\/}, {\em
  D17\/}, 2369--2374, 1978.
\end{thebibliography}

\end{article}
\end{document}